\documentclass[showpacs,twocolumn,amsmath,amssymb,prl,superscriptaddress]{revtex4}

\usepackage[T1]{fontenc}
\usepackage[utf8]{inputenc}
\usepackage[english]{babel}
\usepackage{graphicx}
\usepackage{amsmath,amsfonts,amssymb,amsthm,bm,bbold}
\usepackage{fancyhdr}
\usepackage{mathrsfs}
\usepackage{epstopdf}
\usepackage{color}
\usepackage{nicefrac}
\usepackage{braket}
\usepackage{ulem}

\usepackage{tikz}
\usetikzlibrary{arrows,decorations.pathmorphing,decorations.markings,decorations.shapes,backgrounds,positioning,fit,trees,calc,arrows.meta} 
\tikzset{snake it/.style={decorate, decoration={snake, segment length=1.5mm, amplitude=.4mm}}}
\usepackage{feynmp}

\tikzset{>={Latex[scale=0.45]}}
\tikzset{->-/.style={decoration={
  markings,
  mark=at position #1 with {\arrow{>}}},postaction={decorate}}}
\tikzset{
   vertex/.style={circle, inner sep=0pt, minimum size=5pt,fill=black,label=#1}, vertex/.default=\text{},
   crossing/.style={circle, inner sep=0, minimum size=0,label=#1}, crossing/.default=\text{},
   named/.style={draw,circle, inner sep=0pt, minimum size=12pt},
   namedE/.style={draw,circle, inner sep=0pt, minimum size=7pt},
   bath/.style={draw,thick,->-=.5}, bath/.default={},
   bathr/.style={draw,thick,dashed,->-=.5},
   time/.style={draw,dashed,thin},
   system/.style={draw,thick}, system/.default={},
   syslabel/.style={midway,auto,green!40!black}}
\tikzset{dotted pattern/.style args={#1 and #2}{
   decorate,
   fill=black,
   decoration={
    shape backgrounds,
    shape=circle,
    shape size=#1,
    shape sep={#2, between center},
    }
  },
  dotted pattern/.default={1pt and 1.5mm},
}

\begin{document}

\title{Estimating the error of an analog quantum simulator by additional measurements}

\author{Iris Schwenk}
\affiliation{Institute of Theoretical Solid State Physics (TFP),
      Karlsruhe Institute of Technology (KIT), 76131 Karlsruhe, Germany}

\author{Sebastian Zanker}
\affiliation{Institute of Theoretical Solid State Physics (TFP),
      Karlsruhe Institute of Technology (KIT), 76131 Karlsruhe, Germany}

\author{Jan-Michael Reiner}
\affiliation{Institute of Theoretical Solid State Physics (TFP),
      Karlsruhe Institute of Technology (KIT), 76131 Karlsruhe, Germany}

\author{Juha Lepp\"akangas}
\affiliation{Institute of Theoretical Solid State Physics (TFP),
      Karlsruhe Institute of Technology (KIT), 76131 Karlsruhe, Germany}

\author{Michael Marthaler}
\affiliation{Institut für Theorie der Kondensierten Materie (TKM), Karlsruhe Institute of Technology (KIT), 76131 Karlsruhe, Germany}
\affiliation{Theoretical Physics, Saarland University, 66123 Saarbrücken, Germany}




\date{\today}

\begin{abstract}
We study an analog quantum simulator coupled to a reservoir with a known spectral density.
The reservoir perturbs the quantum simulation by causing decoherence.
The simulator is used to measure an operator average, which cannot be calculated using any classical means.
Since we cannot predict the result, it is difficult to estimate the effect of the environment.
Especially, it is difficult to resolve whether the perturbation is small or if the actual result of the simulation is in fact very different from the ideal system we intend to study. 
Here, we show that in specific systems a measurement of additional correlators can be used to verify the reliability of the quantum simulation.
The procedure only requires additional measurements on the quantum simulator itself.
We demonstrate the method theoretically in the case of a single spin connected to a bosonic environment.
\end{abstract}

\maketitle


In recent years, there has been tremendous progress in the development of qubits for quantum-information applications~\cite{Ion_Review,Rydberg_Review,Martinis_Threshold,Blatt_Perspectives}. 
The number of qubits we need to build a universal error-corrected quantum computer is, however, still relatively large~\cite{Fowler_Error_Correction}. 
Therefore, it is important to develop ideas for near term applications of well-controlled quantum systems. 
One of the key ideas is to use small artificial quantum systems as analog quantum simulators \cite{Analog_Manousakis}.
In this case, we need to be able to map the Hamiltonian of the well-controlled artificial quantum system to a system of interest which we want to simulate~\cite{QS_Review_Georgescu}.
This can either be achieved by rapid application of pulses or by creating a static Hamiltonian which maps to the Hamiltonian of interest, either directly or via a unitary transformation.

Many impressive experiments have been performed recently to examine models from solid-state physics using artificial quantum systems~\cite{Bohnet_219ions,Struck_Frustrated,Strongly_Interacting,Hart_Hubbard} and many additional proposals exist~\cite{Marthaler_hubbard,Tian_Simu_1,Simu_Plenio}.
Even though the mappings are often more complicated, quantum simulation is also applicable to other fields like particle and astrophysics~\cite{Solano_Tachyon,Steinhauer_Hawking}.
However, to use quantum simulation as a tool to investigate problems that are not solvable analytically or on classical computers, it is essential to be able to detect errors in these systems~\cite{Can_we_trust_Emulator,Reliability_AQS,Certification_Eisert}.
Relying on validations using classical computation~\cite{Valdidate_Troyer1,Valdidate_Troyer2} or different physical realizations~\cite{Cross_Validation_Leibfried,Quantum_Simulator_EPJ} strongly restricts the applicability of quantum simulation.

Here, we derive a protocol that allows us to estimate the size of the error in the result of the simulation by additional measurements.
We consider a situation where the error arises from a coupling of the simulator to external degrees of freedom in an environmental bath.
The simulator is described by the Hamiltonian $H_S(t)$ and the goal of the simulation is to determine an operator average $\langle \hat{A}(t)\rangle_0$. 
We discuss a case where the simulator is coupled via the operator $\hat{O}$ to the bath described by the Hamiltonian $H_B$. 
The bath is characterized by the correlator $\langle \hat{X}(t)\hat{X}(0)\rangle_0$, where 
$\hat{X}$ is an operator acting on the bath. 
Many methods exist to determine the bath correlation function~\cite{Bath_Estimation_Bylander,Bath_Estimation_Wilhelm} and we assume that the noise correlator is known. 
This is a simplified model which allows us to explain the major features of our approach. 
However, the approach also works in more general cases where many baths are coupled via many operators $\hat{O}_i$ and in situations where the perturbation is given by disorder in $H_S(t)$. 

Our main result is that an estimate for the error in the simulation can be achieved by measuring additional correlators of the perturbed quantum simulator. 
With these additional measurements it is possible to define a perturbative quantity which 
needs to be small.
To calculate the relevant perturbative quantity it is necessary to measure correlators of the form $\braket{\hat{O}(t_1)\hat{A}(t)\hat{O}(t_2)}$.
Here, we show under which condition the result of the quantum simulation is reliable.
To derive this condition we use an expansion in the coupling on Keldysh contour~\cite{SchoellerSchoenKeldysh} and discuss the conditions of convergence.


The full system can be described by the Hamiltonian
\begin{equation}
H(t) = H_S(t) + H_B + \lambda_B H_C\,, 
\end{equation}
with $\lambda_B=1$ where $H_C$ describes the coupling between the ideal simulator contained in $H_S(t)$ and the bath contained in $H_B$.
The time evolution is given by
\begin{equation} \label{eq_Ideal_Time_Evolution}
U_{\lambda_B}(t,t_0) = {\cal T}\exp\left(-i \int_{t_0}^t \mathrm{d}t' \, H(t')\right) \,,
\end{equation}
where $\cal{T}$ is the time-ordering operator ($t>t_0$).
$U_1(t,t_0)$ describes the time evolution of the full system.
The evolution without coupling to the environment, which we call the ideal time evolution, is described by $U_0(t,t_0)$.
We transform operators in the interaction picture according to,
$\hat{B}_I(t)~=~U_0(t_0,t) \hat{B} U_0(t,t_0)$.
With this we calculate the expectation value of operator $\hat{A}$,
\begin{align}
\braket{\hat{A}(t)} &= \mathrm{Tr} \left[ \rho_0 U(t_0,t) \hat{A} U(t,t_0) \right] \nonumber\\
&=\mathrm{Tr} \left[ \rho_0 U_I(t_0,t) \hat{A}_I(t) U_I(t,t_0) \right] \,,
\end{align}
with $ U_I(t,t_0) = {\cal T} \exp\left(-i \int_{t_0}^t \mathrm{d}t' \, H_{C,I}(t')\right) $.
We expand $\braket{\hat{A}(t)}$ in orders of $H_C$.
In the zeroth order we have $\braket{\hat{A}(t)}\approx\mathrm{Tr} \left[ \rho_0 \hat{A}_I(t) \right]$, which is the ideal result.
To show the expansion in a compact fashion we use the graphical form of an expansion on Keldysh contour~\cite{SchoellerSchoenKeldysh}. In this case an expansion in $H_C$ results in
\begin{eqnarray} \label{eq_diagrammatic_series}
 \langle \hat{A}(t)\rangle  \!=\!\!\!\!\!&&
 \begin{tikzpicture}[anchor=base,baseline=8pt]
    \coordinate (A) at (0,0);
    \coordinate (B) at (0.62,0);
    \coordinate (C) at (0,0.7);
    \coordinate (D) at (0.62,0.7);
    \draw[line width=1.0pt, <-] (A) -- (B);
    \draw[line width=1.0pt, <-] (D) -- (C);
    \end{tikzpicture}
    +
    \begin{tikzpicture}[anchor=base,baseline=8pt]
    \coordinate (A) at (0,0);
    \coordinate (B) at (0.62,0);
    \coordinate (C) at (0,0.7);
    \coordinate (D) at (0.62,0.7);
     \coordinate (A1) at (0.2,0);
    \coordinate (D1) at (0.4,0.7);
    \draw[line width=1.0pt, <-] (A) -- (B);
    \draw[line width=1.0pt, <-] (D) -- (C);
    \fill (A1) circle (2pt);
    \fill (D1) circle (2pt);
    \end{tikzpicture}\!+
     \begin{tikzpicture}[anchor=base,baseline=8pt]
    \coordinate (A) at (0,0);
    \coordinate (B) at (0.62,0);
    \coordinate (C) at (0,0.7);
    \coordinate (D) at (0.62,0.7);
     \coordinate (A1) at (0.4,0);
    \coordinate (D1) at (0.2,0.7);
    \draw[line width=1.0pt, <-] (A) -- (B);
    \draw[line width=1.0pt, <-] (D) -- (C);
    \fill (A1) circle (2pt);
    \fill (D1) circle (2pt);
    \end{tikzpicture}\!+
     \begin{tikzpicture}[anchor=base,baseline=8pt]
    \coordinate (A) at (0,0);
    \coordinate (B) at (0.62,0);
    \coordinate (C) at (0,0.7);
    \coordinate (D) at (0.62,0.7);
     \coordinate (A1) at (0.2,0.7);
    \coordinate (D1) at (0.4,0.7);
    \draw[line width=1.0pt, <-] (A) -- (B);
    \draw[line width=1.0pt, <-] (D) -- (C);
    \fill (A1) circle (2pt);
    \fill (D1) circle (2pt);
    \end{tikzpicture}\!+ 
     \begin{tikzpicture}[anchor=base,baseline=8pt]
    \coordinate (A) at (0,0);
    \coordinate (B) at (0.62,0);
    \coordinate (C) at (0,0.7);
    \coordinate (D) at (0.62,0.7);
     \coordinate (A1) at (0.2,0);
    \coordinate (D1) at (0.4,0);
    \draw[line width=1.0pt, <-] (A) -- (B);
    \draw[line width=1.0pt, <-] (D) -- (C);
    \fill (A1) circle (2pt);
    \fill (D1) circle (2pt);
    \end{tikzpicture}\!\nonumber\\
   &+&\!\!  \begin{tikzpicture}[anchor=base,baseline=8pt]
    \coordinate (A) at (-0.1,0);
    \coordinate (B) at (0.9,0);
    \coordinate (C) at (-0.1,0.7);
    \coordinate (D) at (0.9,0.7);
    \coordinate (A1) at (0.1,0.7);
    \coordinate (B1) at (0.3,0.7);
    \coordinate (C1) at (0.5,0.7);
    \coordinate (D1) at (0.7,0.7);
    \draw[line width=1.0pt, <-] (A) -- (B);
    \draw[line width=1.0pt, <-] (D) -- (C);
    \fill (A1) circle (2pt);
    \fill (B1) circle (2pt);
    \fill (C1) circle (2pt);
    \fill (D1) circle (2pt);
    \end{tikzpicture}\!+
     \begin{tikzpicture}[anchor=base,baseline=8pt]
    \coordinate (A) at (-0.1,0);
    \coordinate (B) at (0.9,0);
    \coordinate (C) at (-0.1,0.7);
    \coordinate (D) at (0.9,0.7);
    \coordinate (A1) at (0.1,0.);
    \coordinate (B1) at (0.3,0.7);
    \coordinate (C1) at (0.5,0.7);
    \coordinate (D1) at (0.7,0.7);
    \draw[line width=1.0pt, <-] (A) -- (B);
    \draw[line width=1.0pt, <-] (D) -- (C);
    \fill (A1) circle (2pt);
    \fill (B1) circle (2pt);
    \fill (C1) circle (2pt);
    \fill (D1) circle (2pt);
    \end{tikzpicture}\!+ 
      \begin{tikzpicture}[anchor=base,baseline=8pt]
    \coordinate (A) at (-0.1,0);
    \coordinate (B) at (0.9,0);
    \coordinate (C) at (-0.1,0.7);
    \coordinate (D) at (0.9,0.7);
    \coordinate (A1) at (0.1,0.);
    \coordinate (B1) at (0.3,0.);
    \coordinate (C1) at (0.5,0.);
    \coordinate (D1) at (0.7,0.7);
    \draw[line width=1.0pt, <-] (A) -- (B);
    \draw[line width=1.0pt, <-] (D) -- (C);
    \fill (A1) circle (2pt);
    \fill (B1) circle (2pt);
    \fill (C1) circle (2pt);
    \fill (D1) circle (2pt);
    \end{tikzpicture}\!+
     \begin{tikzpicture}[anchor=base,baseline=8pt]
    \coordinate (A) at (-0.1,0);
    \coordinate (B) at (0.9,0);
    \coordinate (C) at (-0.1,0.7);
    \coordinate (D) at (0.9,0.7);
    \coordinate (A1) at (0.1,0.);
    \coordinate (B1) at (0.3,0.);
    \coordinate (C1) at (0.5,0.);
    \coordinate (D1) at (0.7,0.);
    \draw[line width=1.0pt, <-] (A) -- (B);
    \draw[line width=1.0pt, <-] (D) -- (C);
    \fill (A1) circle (2pt);
    \fill (B1) circle (2pt);
    \fill (C1) circle (2pt);
    \fill (D1) circle (2pt);
    \end{tikzpicture}\!\nonumber\\
    &+&\!\!  \begin{tikzpicture}[anchor=base,baseline=8pt]
    \coordinate (A) at (-0.1,0);
    \coordinate (B) at (0.9,0);
    \coordinate (C) at (-0.1,0.7);
    \coordinate (D) at (0.9,0.7);
    \coordinate (A1) at (0.1,0.7);
    \coordinate (B1) at (0.3,0.);
    \coordinate (C1) at (0.5,0.7);
    \coordinate (D1) at (0.7,0.7);
    \draw[line width=1.0pt, <-] (A) -- (B);
    \draw[line width=1.0pt, <-] (D) -- (C);
    \fill (A1) circle (2pt);
    \fill (B1) circle (2pt);
    \fill (C1) circle (2pt);
    \fill (D1) circle (2pt);
    \end{tikzpicture}\!+
     \begin{tikzpicture}[anchor=base,baseline=8pt]
    \coordinate (A) at (-0.1,0);
    \coordinate (B) at (0.9,0);
    \coordinate (C) at (-0.1,0.7);
    \coordinate (D) at (0.9,0.7);
    \coordinate (A1) at (0.1,0.7);
    \coordinate (B1) at (0.3,0.7);
    \coordinate (C1) at (0.5,0.);
    \coordinate (D1) at (0.7,0.7);
    \draw[line width=1.0pt, <-] (A) -- (B);
    \draw[line width=1.0pt, <-] (D) -- (C);
    \fill (A1) circle (2pt);
    \fill (B1) circle (2pt);
    \fill (C1) circle (2pt);
    \fill (D1) circle (2pt);
    \end{tikzpicture}\!+ 
      \begin{tikzpicture}[anchor=base,baseline=8pt]
    \coordinate (A) at (-0.1,0);
    \coordinate (B) at (0.9,0);
    \coordinate (C) at (-0.1,0.7);
    \coordinate (D) at (0.9,0.7);
    \coordinate (A1) at (0.1,0.7);
    \coordinate (B1) at (0.3,0.7);
    \coordinate (C1) at (0.5,0.7);
    \coordinate (D1) at (0.7,0.0);
    \draw[line width=1.0pt, <-] (A) -- (B);
    \draw[line width=1.0pt, <-] (D) -- (C);
    \fill (A1) circle (2pt);
    \fill (B1) circle (2pt);
    \fill (C1) circle (2pt);
    \fill (D1) circle (2pt);
    \end{tikzpicture}\!+
     \begin{tikzpicture}[anchor=base,baseline=8pt]
    \coordinate (A) at (-0.1,0);
    \coordinate (B) at (0.9,0);
    \coordinate (C) at (-0.1,0.7);
    \coordinate (D) at (0.9,0.7);
    \coordinate (A1) at (0.1,0.7);
    \coordinate (B1) at (0.3,0.7);
    \coordinate (C1) at (0.5,0.);
    \coordinate (D1) at (0.7,0.);
    \draw[line width=1.0pt, <-] (A) -- (B);
    \draw[line width=1.0pt, <-] (D) -- (C);
    \fill (A1) circle (2pt);
    \fill (B1) circle (2pt);
    \fill (C1) circle (2pt);
    \fill (D1) circle (2pt);
    \end{tikzpicture}\!\nonumber\\
     &+&\!\!  \begin{tikzpicture}[anchor=base,baseline=8pt]
    \coordinate (A) at (-0.1,0);
    \coordinate (B) at (0.9,0);
    \coordinate (C) at (-0.1,0.7);
    \coordinate (D) at (0.9,0.7);
    \coordinate (A1) at (0.1,0.);
    \coordinate (B1) at (0.3,0.);
    \coordinate (C1) at (0.5,0.7);
    \coordinate (D1) at (0.7,0.7);
    \draw[line width=1.0pt, <-] (A) -- (B);
    \draw[line width=1.0pt, <-] (D) -- (C);
    \fill (A1) circle (2pt);
    \fill (B1) circle (2pt);
    \fill (C1) circle (2pt);
    \fill (D1) circle (2pt);
    \end{tikzpicture}\!+
     \begin{tikzpicture}[anchor=base,baseline=8pt]
    \coordinate (A) at (-0.1,0);
    \coordinate (B) at (0.9,0);
    \coordinate (C) at (-0.1,0.7);
    \coordinate (D) at (0.9,0.7);
    \coordinate (A1) at (0.1,0.7);
    \coordinate (B1) at (0.3,0.);
    \coordinate (C1) at (0.5,0.);
    \coordinate (D1) at (0.7,0.);
    \draw[line width=1.0pt, <-] (A) -- (B);
    \draw[line width=1.0pt, <-] (D) -- (C);
    \fill (A1) circle (2pt);
    \fill (B1) circle (2pt);
    \fill (C1) circle (2pt);
    \fill (D1) circle (2pt);
    \end{tikzpicture}\!+ 
      \begin{tikzpicture}[anchor=base,baseline=8pt]
    \coordinate (A) at (-0.1,0);
    \coordinate (B) at (0.9,0);
    \coordinate (C) at (-0.1,0.7);
    \coordinate (D) at (0.9,0.7);
    \coordinate (A1) at (0.1,0.7);
    \coordinate (B1) at (0.3,0.);
    \coordinate (C1) at (0.5,0.7);
    \coordinate (D1) at (0.7,0.);
    \draw[line width=1.0pt, <-] (A) -- (B);
    \draw[line width=1.0pt, <-] (D) -- (C);
    \fill (A1) circle (2pt);
    \fill (B1) circle (2pt);
    \fill (C1) circle (2pt);
    \fill (D1) circle (2pt);
    \end{tikzpicture}\!+
     \begin{tikzpicture}[anchor=base,baseline=8pt]
    \coordinate (A) at (-0.1,0);
    \coordinate (B) at (0.9,0);
    \coordinate (C) at (-0.1,0.7);
    \coordinate (D) at (0.9,0.7);
    \coordinate (A1) at (0.1,0.);
    \coordinate (B1) at (0.3,0.7);
    \coordinate (C1) at (0.5,0.);
    \coordinate (D1) at (0.7,0.7);
    \draw[line width=1.0pt, <-] (A) -- (B);
    \draw[line width=1.0pt, <-] (D) -- (C);
    \fill (A1) circle (2pt);
    \fill (B1) circle (2pt);
    \fill (C1) circle (2pt);
    \fill (D1) circle (2pt);
    \end{tikzpicture}\!\nonumber\\
     &+&\!\!  \begin{tikzpicture}[anchor=base,baseline=8pt]
    \coordinate (A) at (-0.1,0);
    \coordinate (B) at (0.9,0);
    \coordinate (C) at (-0.1,0.7);
    \coordinate (D) at (0.9,0.7);
    \coordinate (A1) at (0.1,0.);
    \coordinate (B1) at (0.3,0.7);
    \coordinate (C1) at (0.5,0.7);
    \coordinate (D1) at (0.7,0.);
    \draw[line width=1.0pt, <-] (A) -- (B);
    \draw[line width=1.0pt, <-] (D) -- (C);
    \fill (A1) circle (2pt);
    \fill (B1) circle (2pt);
    \fill (C1) circle (2pt);
    \fill (D1) circle (2pt);
    \end{tikzpicture}\!+
     \begin{tikzpicture}[anchor=base,baseline=8pt]
    \coordinate (A) at (-0.1,0);
    \coordinate (B) at (0.9,0);
    \coordinate (C) at (-0.1,0.7);
    \coordinate (D) at (0.9,0.7);
    \coordinate (A1) at (0.1,0.7);
    \coordinate (B1) at (0.3,0.);
    \coordinate (C1) at (0.5,0.);
    \coordinate (D1) at (0.7,0.7);
    \draw[line width=1.0pt, <-] (A) -- (B);
    \draw[line width=1.0pt, <-] (D) -- (C);
    \fill (A1) circle (2pt);
    \fill (B1) circle (2pt);
    \fill (C1) circle (2pt);
    \fill (D1) circle (2pt);
    \end{tikzpicture}\!+ 
      \begin{tikzpicture}[anchor=base,baseline=8pt]
    \coordinate (A) at (-0.1,0);
    \coordinate (B) at (0.9,0);
    \coordinate (C) at (-0.1,0.7);
    \coordinate (D) at (0.9,0.7);
    \coordinate (A1) at (0.1,0.);
    \coordinate (B1) at (0.3,0.7);
    \coordinate (C1) at (0.5,0.);
    \coordinate (D1) at (0.7,0.);
    \draw[line width=1.0pt, <-] (A) -- (B);
    \draw[line width=1.0pt, <-] (D) -- (C);
    \fill (A1) circle (2pt);
    \fill (B1) circle (2pt);
    \fill (C1) circle (2pt);
    \fill (D1) circle (2pt);
    \end{tikzpicture}\!+
     \begin{tikzpicture}[anchor=base,baseline=8pt]
    \coordinate (A) at (-0.1,0);
    \coordinate (B) at (0.9,0);
    \coordinate (C) at (-0.1,0.7);
    \coordinate (D) at (0.9,0.7);
    \coordinate (A1) at (0.1,0.);
    \coordinate (B1) at (0.3,0.);
    \coordinate (C1) at (0.5,0.7);
    \coordinate (D1) at (0.7,0.);
    \draw[line width=1.0pt, <-] (A) -- (B);
    \draw[line width=1.0pt, <-] (D) -- (C);
    \fill (A1) circle (2pt);
    \fill (B1) circle (2pt);
    \fill (C1) circle (2pt);
    \fill (D1) circle (2pt);
    \end{tikzpicture}\! +\dots \,.
\end{eqnarray}
Here, each upper line corresponds to the free (ideal) forward time evolution, Eq.~(\ref{eq_Ideal_Time_Evolution}) with $\lambda_B=0$, and the lower line to the backward time evolution.
The dots represent orders of $H_C$ and we consider explicit time sorting.
We first discuss contributions up to second order and later we confirm that a small perturbation in second order indicates a small contribution to all orders by analyzing the fourth order.
All diagrams up to the fourth order are shown.
We assume $\braket{H_C}=0$ and that contributions from odd orders in $H_C$ are zero.
Explicitly, the contributions up to second order are
\begin{equation} 
\braket{A(t)} \approx\mathrm{Tr} \left[ \rho_0 \hat{A}_I(t) \right] +\int\limits_{t_0}^{t}\!\mathrm{d}t_1\!\int\limits_{t_0}^{t_1}\!\mathrm{d}t_2\, C_2(t,t_1,t_2) \,,
\end{equation}
with the zeroth-order term $\mathrm{Tr} \left[ \rho_0 \hat{A}_I(t) \right]=\braket{\hat{A}(t)}_0$, where the index $0$ indicates that the time evolution is taken with $U_{\lambda_B=0}$.
If we assume a very general form for the coupling Hamiltonian $H_C = \hat{O}\hat{X}$ (with $ \hat{O}\hat{X}= \hat{O}\otimes\hat{X}$), the lowest-order correction $C_2(t,t_1,t_2)$ can be written as,
\begin{align}\label{eq_C2_special}
C_2(t,t_1,t_2) =&\braket{\hat{O}(t_1)\hat{A}(t) \hat{O}(t_2)}_0\braket{\hat{X}(t_1)\hat{X}(t_2)}_0 \nonumber\\
&+\braket{\hat{O}(t_2)\hat{A}(t) \hat{O}(t_1)}_0\braket{\hat{X}(t_2)\hat{X}(t_1)}_0 \nonumber \\
&-\braket{\hat{O}(t_2)\hat{O}(t_1)\hat{A}(t) }_0\braket{\hat{X}(t_2)\hat{X}(t_1)}_0 \nonumber \\
&-\braket{\hat{A}(t)\hat{O}(t_1)\hat{O}(t_2)}_0\braket{\hat{X}(t_1)\hat{X}(t_2)}_0  \,.
\end{align}
If the expansion converges, $C_2$ is an estimate for the size of the error introduced by the environment.
This is discussed in more detail at the end of this article.
To be able to estimate $C_2$, the approximate size and form of the noise correlator $\braket{\hat{X}(t_1)\hat{X}(t_2)}_0$ needs to be known.
We assume that the noise correlator can be estimated either by theoretical considerations or through calibration techniques. 

We now propose the following protocol to estimate the deviation of the measurement result $\langle \hat{A}(t)\rangle$ from the ideal result $\langle \hat{A}(t)\rangle_0$:
\begin{enumerate}
 \item The average $\langle \hat{A}(t)\rangle$ and the three-time correlators of the form $\braket{\hat{O}(t_1)\hat{A}(t) \hat{O}(t_2)}$ which appear in Eq.~(\ref{eq_C2_special}) have to be measured.
Of course, when measured, these quantities are perturbed averages with $H_C\neq 0 $.
 \item We assume the perturbation to be small, an assumption which we will control in a self-consistent way.
From our initial assumption we can then assume that $\langle \hat{A}(t)\rangle\approx \langle \hat{A}(t)\rangle_0$.
In the same way we assume for all relevant three-time correlators that 
 $\braket{\hat{O}(t_1)\hat{A}(t) \hat{O}(t_2)}\approx \braket{\hat{O}(t_1)\hat{A}(t) \hat{O}(t_2)}_0$ and so on for the other correlators.
 \item In accordance with our assumption in point 2, we use the measured three-time correlators to calculate $C_2$. 
If all our assumptions are valid, the correction is small in comparison to the result of the simulation $\braket{\hat{A}(t)}$ and we have a relation of the form
\begin{equation} \label{eq_condition_C_2<A}
\left|\int\limits_{t_0}^{t}\!\mathrm{d}t_1\!\int\limits_{t_0}^{t_1}\!\mathrm{d}t_2\, C_2(t,t_1,t_2) \right|\ll \left|\braket{\hat{A}(t)}\right| \,.
\end{equation}
\end{enumerate}
If Eq.~(\ref{eq_condition_C_2<A}) holds for the $C_2$ calculated from the measurement results on the perturbed simulator, this means that the determination of $C_2$ was approximately correct and that the results of the quantum simulator are reliable.
In the case that Eq.~(\ref{eq_condition_C_2<A}) does not hold, both our estimate for $C_2$ and for $\braket{\hat{A}(t)}_0$ are not reliable and within our method no result can be gained from the simulator.

To carry out the protocol correlators of three operators at different times have to be measured.
Techniques to measure two-time correlators\cite{correlators_Laflamme, correlators_Troyer, correlators_FWM} using only a single ancilla qubit, can be extended to the case of three-time correlators.
However, in the following we describe how the reliability can be estimated without the need of three-time correlators and the exact knowledge of the bath.

For any situation where the effect of the bath should stay small, it is a necessary assumption to approximate the bath correlation function with an exponential decay:
\begin{equation}\label{eq_Assumptions_for_XX_correlator}
\braket{\hat{X}(t_1)\hat{X}(t_2)}_0 \approx \lambda e^{-\gamma|t_1-t_2|} \,.
\end{equation} 
It is also in accordance with the standard master-equation approach used to describe decoherence in quantum systems.
In a more general treatment, the bath correlator is given by a sum of exponential functions.
In our approximation, we only keep the function with the smallest decay rate $\gamma$ which contributes at long time scales.
With Eq.~(\ref{eq_Assumptions_for_XX_correlator}) we find,
\begin{equation}
\int\limits_{t_0}^{t}\!\mathrm{d}t_1\!\int\limits_{t_0}^{t_1}\!\mathrm{d}t_2\,|\braket{\hat{X}(t_1)\hat{X}(t_2)}_0| \approx \frac{\lambda}{\gamma} (t-t_0) \,.
\end{equation}
An upper bound for $C_2$ is then given by
\begin{align}
\left|\int\limits_{t_0}^{t}\!\mathrm{d}t_1\!\int\limits_{t_0}^{t_1}\!\mathrm{d}t_2\, C_2(t,t_1,t_2) \right|\leq& \frac{\lambda}{\gamma} (t-t_0) \left(2 |\braket{\hat{O}(t)\hat{A}(t)\hat{O}(t)}_0|\right.\nonumber\\
&+|\braket{\hat{A}(t)\hat{O}(t)\hat{O}(t)}_0| \nonumber\\
&\left.+|\braket{\hat{O}(t)\hat{O}(t)\hat{A}(t)}_0| \right) \,,
\label{eq_simple_condition_C_2<A}
\end{align}
where the system correlators are estimated by their maximal value.
Using this result for the left-hand side of Eq.~(\ref{eq_condition_C_2<A}) it is possible to verify the reliability of the simulation, without measuring three-time correlators.
With this we see that a measurement of system correlators at one time and a rough knowledge of the decay properties of the bath is enough to estimate if the result of the simulation is trustworthy.


Using the protocol described above it is possible to evaluate for which time $t$ the result of the simulator is reliable. But under specific circumstances the corrections even remain small for all times.
It is immediately clear that our expansion corresponds to a short-term expansion of the time evolution.
Therefore, our expansion tends to diverge in the limit of long times for generic three-time correlators $\braket{\hat{O}(t_1)\hat{A}(t) \hat{O}(t_2)}_0$.
This means that for most simulations the results will be reliable on a certain time scale.
Using our protocol it is possible to evaluate what this time scale is.
However, we find that for some specific but still quite general forms of the three-time correlators, convergence can be achieved even in the long time limit and therefore the error in a quantum simulation might stay small on all time scales.
After this, we verify for the case of a single qubit that the protocol described above is in fact able to characterize the reliability of the result of the perturbed simulation.
In the following, we call the three-time correlators $\braket{\hat{O}(t_1)\hat{A}(t) \hat{O}(t_2)}_0$ "system correlators" since they only involve operators acting on the system.

To analyze the convergence of the series in Eq.~(\ref{eq_diagrammatic_series}),
we approximate the bath correlator as an exponentially decaying function (Eq.~(\ref{eq_Assumptions_for_XX_correlator})) and assume the bath to be Markovian. This means that the smallest decay rate $\gamma$ is still large in comparison to the system decay rates. For good quantum information systems this assumption is reasonable.

At first, we consider the worst case scenario, where the system correlators neither decay nor oscillate fast.
Here, we treat them as a constant factor $c_1$ when evaluating the integrals.
With this we find,
\begin{align}
&\int\limits_{t_0}^{t}\!\mathrm{d}t_1\!\int\limits_{t_0}^{t_1}\!\mathrm{d}t_2 \braket{\hat{O}(t_1)\hat{A}(t) \hat{O}(t_2)}_0 \braket{X(t_1)X(t_2)}_0\nonumber \\ 
&= c_1 \lambda \frac{t-t_0}{\gamma} + c_1 \lambda \frac{e^{-\gamma(t-t_0)}-1}{\gamma^2} \, .
\end{align}
For each of the four terms in Eq.~(\ref{eq_C2_special}) we find the same result with an appropriate constant $c_i$.
These terms increase with increasing time.
In this case the series converges only in the short-time limit. Consequently, the results of the simulation are only reliable on the respective time scale.
It is illustrative to consider a term of the fourth order.
Here, we contract the bath correlators in the following form
\begin{equation}
\begin{tikzpicture}[anchor=base,baseline=8pt]
    \coordinate (A) at (-0.1,0);
    \coordinate (B) at (0.9,0);
    \coordinate (C) at (-0.1,0.7);
    \coordinate (D) at (0.9,0.7);
    \coordinate (A1) at (0.1,0.);
    \coordinate (B1) at (0.3,0.7);
    \coordinate (C1) at (0.5,0.);
    \coordinate (D1) at (0.7,0.7);
    \draw[line width=1.0pt, <-] (A) -- (B);
    \draw[line width=1.0pt, <-] (D) -- (C);
    \fill (A1) circle (2pt);
    \fill (B1) circle (2pt);
    \fill (C1) circle (2pt);
    \fill (D1) circle (2pt);
    \end{tikzpicture}
= 
\begin{tikzpicture}[anchor=base,baseline=8pt]
    \coordinate (A) at (-0.1,0);
    \coordinate (B) at (0.9,0);
    \coordinate (C) at (-0.1,0.7);
    \coordinate (D) at (0.9,0.7);
    \coordinate (A1) at (0.1,0.);
    \coordinate (B1) at (0.3,0.7);
    \coordinate (C1) at (0.5,0.);
    \coordinate (D1) at (0.7,0.7);
    \draw[line width=1.0pt, <-] (A) -- (B);
    \draw[line width=1.0pt, <-] (D) -- (C);
    \fill (A1) circle (2pt);
    \fill (B1) circle (2pt);
    \fill (C1) circle (2pt);
    \fill (D1) circle (2pt);
    \draw[line width=1.0pt,snake it] (A1) -- (B1);
    \draw[line width=1.0pt,snake it] (C1) -- (D1);
    \end{tikzpicture}
+
\begin{tikzpicture}[anchor=base,baseline=8pt]
    \coordinate (A) at (-0.1,0);
    \coordinate (B) at (0.9,0);
    \coordinate (C) at (-0.1,0.7);
    \coordinate (D) at (0.9,0.7);
    \coordinate (A1) at (0.1,0.);
    \coordinate (B1) at (0.3,0.7);
    \coordinate (C1) at (0.5,0.);
    \coordinate (D1) at (0.7,0.7);
    \draw[line width=1.0pt, <-] (A) -- (B);
    \draw[line width=1.0pt, <-] (D) -- (C);
    \fill (A1) circle (2pt);
    \fill (B1) circle (2pt);
    \fill (C1) circle (2pt);
    \fill (D1) circle (2pt);
    \draw[line width=1.0pt,snake it] (A1) -- (D1);
    \draw[line width=1.0pt,snake it] (C1) -- (B1);
    \end{tikzpicture}
+
\begin{tikzpicture}[anchor=base,baseline=8pt]
    \coordinate (A) at (-0.1,0);
    \coordinate (B) at (0.9,0);
    \coordinate (C) at (-0.1,0.7);
    \coordinate (D) at (0.9,0.7);
    \coordinate (A1) at (0.1,0.);
    \coordinate (B1) at (0.3,0.7);
    \coordinate (C1) at (0.5,0.);
    \coordinate (D1) at (0.7,0.7);
    \draw[line width=1.0pt, <-] (A) -- (B);
    \draw[line width=1.0pt, <-] (D) -- (C);
    \fill (A1) circle (2pt);
    \fill (B1) circle (2pt);
    \fill (C1) circle (2pt);
    \fill (D1) circle (2pt);
    \draw[line width=1.0pt,snake it] (C1) to[out=110,in=70] (A1);
    \draw[line width=1.0pt,snake it] (B1) to[in=-110,out=-70] (D1);
    \end{tikzpicture} \,.
\end{equation}
To write it in this way we assumed that Wick's theorem holds for the operators $\hat{X}$, so that the bath correlator is given by a product of two-time correlators.
From standard master-equation calculations, we find that inseparable diagrams (the last two diagrams) will converge quite well if the bath correlator decays fast. 
For our specific calculation this means that inseparable diagrams will never be larger then the lowest-order contribution. 
However, the first diagram is separable and therefore not converging for $t_0\rightarrow -\infty$,
\begin{align}
&\begin{tikzpicture}[anchor=base,baseline=8pt]
    \coordinate (A) at (-0.1,0);
    \coordinate (B) at (0.9,0);
    \coordinate (C) at (-0.1,0.7);
    \coordinate (D) at (0.9,0.7);
    \coordinate (A1) at (0.1,0.);
    \coordinate (B1) at (0.3,0.7);
    \coordinate (C1) at (0.5,0.);
    \coordinate (D1) at (0.7,0.7);
    \draw[line width=1.0pt, <-] (A) -- (B);
    \draw[line width=1.0pt, <-] (D) -- (C);
    \fill (A1) circle (2pt);
    \fill (B1) circle (2pt);
    \fill (C1) circle (2pt);
    \fill (D1) circle (2pt);
    \draw[line width=1.0pt,snake it] (A1) -- (B1);
    \draw[line width=1.0pt,snake it] (C1) -- (D1);
    \end{tikzpicture} \!=
\!\!\!\int\limits_{t_0}^{t}\!\mathrm{d}t_1\!\!\int\limits_{t_0}^{t_1}\!\!\mathrm{d}t_2\!\!\int\limits_{t_0}^{t_2}\!\!\mathrm{d}t_3\!\!\int\limits_{t_0}^{t_3}\!\!\mathrm{d}t_4 \! \braket{\hat{X}(t_1)X(t_2)}_0 \! \braket{\hat{X}(t_3)X(t_4)}_0 \nonumber \\
&\times \braket{\hat{O}(t_3) \hat{O}(t_1) \hat{A}(t) \hat{O}(t_2) \hat{O}(t_4)}_0 \approx c \lambda^2 \frac{(t-t_0)^2}{\gamma^2} \,.
\end{align}
We see that some of the fourth-order contributions grow quadratically with time and similar results can be found for higher-order diagrams.
To achieve convergence in this series at long times, it is necessary that the correlators of the system operators decay exponentially.

We now demonstrate the convergence in a simple example, where $H_S$ is a simulator for the spin-boson model, a situation which is similar to several proposals for quantum simulators consisting of qubits coupled to bosonic baths \cite{Solano_Bath,Guzik_Bath}.
We consider a regime where all properties of the model can be calculated within the Born-Markov approximation~\cite{Buch_Weiss}.
This, of course, is not a very interesting limit for quantum simulation, but it allows us to straightforwardly discuss the properties of the correction $C_2$, Eq.~(\ref{eq_C2_special}), for the case where the system correlators decay exponentially to the stationary value.
The system we consider is described by
\begin{align}
H_S &=\frac{1}{2} \epsilon \sigma_z + \sum_i \omega_i b_i^{\dagger} b_i + \sigma_x \sum_i t_i (b_i^{\dagger} + b_i) \\
H_B &= \sum_i \epsilon_i c_i^{\dagger} c_i \\
H_C &= \sigma_x \sum_i f_i (c_i^{\dagger} + c_i) \,.
\end{align}
We have then $\hat{O}=\sigma_x$.
Experimentally this could correspond to a situation where a simulator of the spin-boson model
is coupled to another independent bosonic bath.
This offset to the ideal system can be described by the perturbation $H_B$ coupled to the system via $H_C$.
For simplicity, we assume now that the bath $H_B$ has approximately a flat spectral density, so that we can use the simple approximation in Eq.~(\ref{eq_Assumptions_for_XX_correlator}) for the bath correlator (similar exponential decay also characterizes more general environments in the long time limit).
Additionally, we assume the variation of the system correlators $\braket{\hat{O}(t_1)\hat{A}(t) \hat{O}(t_2)}_0$ on the scale $\gamma^{-1}$ to be negligible and therefore in the limit $t_0\rightarrow-\infty$ we have
\begin{align}
&\int\limits_{t_0}^{t}\!\mathrm{d}t_1\!\int\limits_{t_0}^{t_1}\!\mathrm{d}t_2\, C_2(t,t_1,t_2) \nonumber\\
\approx& \frac{2 \lambda}{\gamma} \int\limits_{t_0}^{t}\!\mathrm{d}t_1  \left( \braket{\sigma_{x}(t_1)\hat{A}(t)\sigma_{x}(t_1)}_0 -\braket{\hat{A}(t)}_0 \right) \\
=& \frac{2\lambda}{\gamma} \int\limits_{t_0}^{t}\!\mathrm{d}t_1  \left(\mathrm{Tr}\left[ \Pi^0_{t_1\rightarrow t} (\sigma_x \Pi^0_{t_0\rightarrow t_1} (\rho_0) \sigma_x) \hat{A} \right] \right. \nonumber\\
&\qquad\left.-\mathrm{Tr}\left[ \Pi^0_{t_0\rightarrow t}(\rho_0) \hat{A} \right]\right)\,, \label{eq_C_2_with_sigma_x}
\end{align}
where $\Pi^0_{t_1\rightarrow t_2} (\rho)$ describes the time evolution of the system density matrix according to $H_0$.
To derive this we have used the quantum regression theorem and that $\sigma_x^2=\mathbb{1}$.
Master-equation calculations yield the time evolution of the density matrix of the spin-boson model,
\begin{align}
\Pi^0_{t_0\rightarrow t} (\rho) = \begin{pmatrix}
                                        e^{-\Gamma (t-t_0)} \rho_0^{\uparrow\uparrow} & e^{-\frac{i\epsilon+\Gamma}{2}(t-t_0)}\rho_0^{\uparrow\downarrow}\\
                                        e^{\frac{i\epsilon-\Gamma}{2}(t-t_0)}\rho_0^{\downarrow\uparrow} & 1-e^{-\Gamma (t-t_0)} \rho_0^{\uparrow\uparrow}
                                  \end{pmatrix}\,,
\end{align}
with the spin decay rate $\Gamma$, and $\rho^0$ being the initial density matrix at $t_0$.
Preparing the system in a mixed state,
\begin{equation}\label{eq:mixed_state}
\rho_0=\begin{pmatrix} a & 0 \\ 0 & 1-a \end{pmatrix}\,,
\end{equation}
we find in the limit $t_0\rightarrow-\infty$
\begin{align}
\Pi^0_{t_1\rightarrow t} &(\sigma_x \Pi^0_{t_0\rightarrow t_1} (\rho_0) \sigma_x) - \Pi^0_{t_0\rightarrow t}(\rho_0) \nonumber \\
                       &= \begin{pmatrix}
                                       e^{-\Gamma (t-t_1)}& 0 \\
                                       0 & -e^{-\Gamma (t-t_1)}
                          \end{pmatrix} \,
\end{align}
and with this
\begin{equation}
\int\limits_{t_0}^{t}\!\mathrm{d}t_1\!\int\limits_{t_0}^{t_1}\!\mathrm{d}t_2\, C_2(t,t_1,t_2) \propto \frac{\lambda}{\gamma\Gamma} \,.
\end{equation}
We find a finite value for the second-order term even on long time scales.
With this we see that the system correlators have the ability to create convergence in the series in Eq.~(\ref {eq_diagrammatic_series}) in the long-time limit.
On the condition that the series converges, a measurement of $C_2$ reveals the reliability of the quantum simulation for all times.
In the above situation, the result is then proportional to the ratio of the decoherence rates induced by the two parts of the bosonic environment, $\lambda/\gamma$ and $\Gamma$, as expected.
In the Supplementary Information, we go through another example where $C_2$ correctly predicts the induced error.

To study the question whether a small value of $C_2$ justifies the assumption that the series converges, we use the fact that the main contributions for $t_0 \rightarrow -\infty$ arise from separable diagrams.
For convergence, it is necessary that the system correlators decay.
For a simple example we assume them to decay exponentially like $e^{-\kappa(t-t_i)}$.
With this we find in the second order (similar to above)
\begin{align}
\int\limits_{t_0}^{t}\!\mathrm{d}t_1\!\!&\int\limits_{t_0}^{t_1}\!\!\mathrm{d}t_2 C_2(t,t_1,t_2) \approx \frac{\lambda}{\gamma\kappa}\left(2\braket{\hat{O}(t)\hat{A}(t)\hat{O}(t)}_0 \right. \nonumber\\
&-\left.\braket{\hat{O}(t)\hat{O}(t)\hat{A}(t)}_0-\braket{\hat{A}(t)\hat{O}(t)\hat{O}(t)}_0\right) \,.
\end{align}
In the fourth order, we get the contributions
\begin{align}
&\int\limits_{t_0}^{t}\!\mathrm{d}t_1\!\!\int\limits_{t_0}^{t_1}\!\!\mathrm{d}t_2\int\limits_{t_0}^{t_2}\!\!\mathrm{d}t_3\!\!\int\limits_{t_0}^{t_3}\!\!\mathrm{d}t_4 C_4(t,t_1,t_2,t_3,t_4) \nonumber\\
\approx& \left(\frac{\lambda}{\gamma\kappa}\right)^2 \left( \braket{\hat{O}(t)\hat{O}(t)\hat{O}(t)\hat{O}(t)\hat{A}(t)}_0 \right.\nonumber\\
&\qquad-4\braket{\hat{O}(t)\hat{O}(t)\hat{O}(t)\hat{A}(t)\hat{O}(t)}_0 \nonumber\\
&\qquad+6\braket{\hat{O}(t)\hat{O}(t)\hat{A}(t)\hat{O}(t)\hat{O}(t)}_0 \nonumber\\
&\qquad-4\braket{\hat{O}(t)\hat{A}(t)\hat{O}(t)\hat{O}(t)\hat{O}(t)}_0 \nonumber\\
&\qquad+\left.\braket{\hat{A}(t)\hat{O}(t)\hat{O}(t)\hat{O}(t)\hat{O}(t)}_0 \right) \,.
\end{align}
In the specific case $\hat{O}=\sigma_x$ and $\hat{A}=\sigma_z$, we obtain in second order that
$\int_{t_0}^{t}\!{\rm d}t_1\!\!\int_{t_0}^{t_1}\!\!{\rm d}t_2 C_2(t,t_1,t_2) \approx -4 \frac{\lambda}{\gamma\kappa} \braket{\hat{A}(t)}_0$,
and in fourth order that
\begin{align}
\int\limits_{t_0}^{t}\!\mathrm{d}t_1\!\!\int\limits_{t_0}^{t_1}\!\!\mathrm{d}t_2\int\limits_{t_0}^{t_2}\!\!\mathrm{d}t_3\!\!\int\limits_{t_0}^{t_3}\!\!\mathrm{d}t_4 C_4(t,t_1,t_2,t_3,t_4) &\approx 16 \! \left(\!\frac{\lambda}{\gamma\kappa}\!\right)^2 \!\!\braket{\hat{A}(t)}_0 \nonumber \\
\qquad\approx \left( \int\limits_{t_0}^{t}\!\mathrm{d}t_1\!\!\int\limits_{t_0}^{t_1}\!\!\mathrm{d}t_2 C_2(t,t_1,t_2) \right)^2& \braket{\hat{A}(t)}_0^{-1}
\end{align}
We see that if $C_2/\braket{A(t)}_0 =x$ then $C_4/\braket{A(t)}_0\approx x^2$.
Therefore, in this example, a small value of $C_2$ indicates the convergence of the series in $H_C$.
In the considered situations a small value of $C_2$ indicated negligible higher-order contributions and thereby justified the verification method.
This is not true in all generality.
Of course, in certain systems it might be possible that $C_4$  is finite even though $C_2$ vanishes, since matrix elements of operators $\hat{A}$ and $\hat{O}$ force that (see Supplementary Information).
With this in mind, the protocol can also be applied to a set of operators $\hat{A}_i$, rather than one, to increase the reliability.
For non-ideal situations beyond these scenarios, further error tests can be considered by extending the above protocol to higher orders in $H_C$.


To use our protocol one has to estimate which operator $\hat{O}$ couples the system to its environment. For the different qubit types the dominating noise sources are well known. Carrying out the protocol for the possible operators $\hat{O}$ separatly will lead to an estimation for the reliability.
It is also possible that the quantum simulator is not coupled to its environment through one operator $\hat{O}$, but rather that multiple independent baths are coupled via $H_C=\sum_i \hat{O}_i \hat{X}_i$.
The calculations presented above can be straightforwardly extended to this case, where sums over $\braket{\hat{O}_{i}(t_1)\hat{A}(t) \hat{O}_{j}(t_2)}_0$ appear.
This is for example the case for a system of multiple qubits, where each qubit is coupled to an individual bath.
Therefore, the protocol scales with $N^2$, where $N$ is the number of qubits.
For large number of qubits the effort to measure the system correlators will be significant.
Because of this, we think our protocol is especially useful for analog simulators consisting of qubits coupled to bosonic modes.
These simulators can yield interesting results using a small number of qubits\cite{Guzik_Bath}.

We derived our protocol on the basis of a time-dependent system Hamiltonian $H_S(t)$.
Formally, this means that the case of digital quantum simulation with a sequence of gates is included in this discussion.
Digital quantum simulation allows us to simulate a great variety of problems \cite{DQS_cQED, DQS_ferm, DQS_ions} but a large number of qubits and gates is necessary for these simulations. Using digital-analog approaches\cite{Solano_Bath} the number of qubits can be reduced, so that our protocol to estimate the reliability is practically relevant.

Furthermore, errors in quantum simulation can arise from imperfections in the fabrication, which can be interpreted as disorder.
In this case, the perturbed quantum simulator is described by $H=H_S+\delta H_S$.
Standard perturbation theory tells us that in case of small static errors the lowest order correction is proportional to matrix elements of $\delta H_S$ and stays small at all times.
For slowly time-dependent disorder $\delta H_S(t)$, it is very likely that the ensemble averages have a Gaussian distribution. With this we end up with equations similar to those discussed above.
For fast fluctuating disorder we get the same equations as above where the bath expectation values are averages of classical random variables.
With this we see that in the case of disorder it is also possible to estimate the reliability of the simulator using such a self-consistent protocol.


In conclusion, we showed how the measurement of additional correlators can be used to study the reliability of quantum simulation of a certain expectation value $\braket{\hat{A}(t)}$ in a self-consistent manner.
The procedure allows to estimate the size of the error by using only measurable quantities of the quantum simulator.
For this purpose, it is necessary to have a rough estimation of the decay properties of the bath.
The protocol then reveals on which time scale the result for $\braket{\hat{A}(t)}$ is reliable. Additionally, we show that for specific systems, where the system correlators decay towards a stationary average, the results of the simulation are reliable for all times.
This type of approach can then be used to test the reliability of the result of an analog quantum simulation without comparing to classical computations or to other physical realizations of the quantum simulator.
Our method supports the development of near term quantum simulators that are used to solve problems which cannot be solved or verified using classical computers.

\begin{acknowledgements}
We thank Gerd Sch\"on for  fruitful discussions.
Iris Schwenk acknowledges financial support by Friedrich-Ebert-Stiftung.
\end{acknowledgements}

\begin{widetext}
\appendix
\section{Supplementary Information}
In this supplementary part, we show that for a system Hamiltonian
that describes a decoherence-free single spin, and a bath Hamiltonian that induces system decoherence, the check of the condition in Eq.~(\ref{eq_condition_C_2<A}) results in the conclusion that the quantum simulation ist not reliable.
With this we have then given examples to show that our protocol works in confirming a reliable quantum simulation.
But as well we showed that the protocol is able to reveal the fact that perturbations are too strong to trust the results.
The case of a single qubit is not of interest for real quantum simulations, but due to the fact that the system correlators do not decay without influence of the external degrees of freedom, it is a simple example to illustrate the rejection by our protocol.

We consider the case, where
\begin{align}
H_S =\frac{1}{2} \epsilon \sigma_z \qquad
H_B = \sum_i \epsilon_i c_i^{\dagger} c_i \qquad
H_C = \sigma_x \sum_i f_i (c_i^{\dagger} + c_i) \,.
\end{align}
The system Hamiltonian is thereby decoherence free.
In our protocol, $C_2$ is calculated (measured) using the perturbed correlators, since we assume self-consistently that they are equivalent (or very close) to the unperturbed correlators.
To show that our protocol works, we have to verify the condition in Eq.~(\ref{eq_condition_C_2<A}) using the perturbed quantities. 

Since the perturbed system is equivalent to the unperturbed spin-boson model discussed in the paper, we can give the result directly,
\begin{equation}
\int\limits_{t_0}^{t}\!\mathrm{d}t_1\!\int\limits_{t_0}^{t_1}\!\mathrm{d}t_2\, C_2(t,t_1,t_2) \propto \frac{\tilde{\lambda}}{\tilde{\gamma}\tilde{\Gamma}} \,,
\end{equation}
where we assume the limit $t_0\rightarrow -\infty$ and used the initial mixed state in Eq.~(\ref{eq:mixed_state}).
The bath that induces the transition of the qubit with rate $\tilde{\Gamma}$, is in this case the same that characterizes the external bath, so that we have $\tilde{\Gamma}\approx\frac{\tilde{\lambda}}{\tilde{\gamma}}$.
Within this identification we get
\begin{equation}
\int\limits_{t_0}^{t}\!\mathrm{d}t_1\!\int\limits_{t_0}^{t_1}\!\mathrm{d}t_2\, C_2(t,t_1,t_2) \approx 2 (A_{\downarrow\downarrow}-A_{\uparrow\uparrow}) \,,
\end{equation}
where $A_{\uparrow\uparrow}$, $A_{\downarrow\downarrow}$ are matrix elements of $\hat{A}$.
In the limit $t_0\rightarrow -\infty$, the condition implied by the relation of Eq.~(\ref{eq_condition_C_2<A}) then reduces to
\begin{equation}
2 \vert A_{\downarrow\downarrow}-A_{\uparrow\uparrow} \vert \ll \vert A_{\downarrow\downarrow} \vert \,.
\end{equation}
In the case $A_{\uparrow\uparrow}=-A_{\downarrow\downarrow}$ (i.e.~when $\hat{A}$ is $\sigma_z$), the condition is never fulfilled.
Therefore, the protocol reveals that the result of the simulation is not reliable.
We also see that applying our protocol to different operators $\hat{A}$ erases the possible mistake arising from that the reliability condition is fulfilled "by accident" (i.e.~when $A_{\uparrow\uparrow}\approx A_{\downarrow\downarrow}$). In this example this would be the case for $\hat{A}$ in the $x$-$y$-plane. Choosing an operator with a contribution of $\sigma_z$ would uncover this false assessment.
\end{widetext}

\end{document}